# A New Keyboard for the Bohlen-Pierce Scale

Antonio B. Nassar

Science Department
The Harvard-Westlake School
3700 Coldwater Canyon
N. Hollywood, CA 91604
And
Department of Math and Sciences
UCLA Extension
10995 Le Conte Avenue
Los Angeles, CA 90024

**Abstract:**

The study of harmonic scales of musical instruments is discussed in all introductory physics texts devoted to the science of sound. In this paper, we present a new piano keyboard to make the so-called Bohlen-Pierce scale more functional and pleasing for composition and performance.

**PACS: 01.55.+b; 01.90.+g**

## 1. Introduction

The basis of harmonic scales of musical instruments is the following:[1,2] If two tones with frequencies $f_1$ and $f_2$ are played together, the result is pleasant to the human ear if the ratio $f_1:f_2$ is equal to m:n, with m and n two integers. This discovery is attributed to the ancient Chinese and Greeks. A series of bells has been discovered in the tomb of the Count of Chin (around 200 B.C.). They were found to be tuned quite precisely to the harmonic scale. On the other hand, the Greeks, with their abundance of string instruments, discovered that dividing a string into equal parts or chopping of one-third of the string resulted in pleasant musical intervals: An octave (2:1) and a perfect fifth (3:2), respectively.

The Pythagoreans asked themselves whether an integral number of octaves could be constructed from the fifth alone by repeated application of the simple frequency ratio 3:2. In mathematical notation, they asked for a solution to[3]

$$\left(\frac{3}{2}\right)^n = 2^m$$

in positive integers n and m. An excellent approximation solution found by the Greeks was:

$$\left(\frac{3}{2}\right)^{12} \approx 2^7.$$

The exponents 12 and 7 are co-prime, so that repeated application of the perfect fifth modulo the octave (the circle of fifths) will not be close to a previously generated frequency until the twelfth step.[3] These 12 different frequencies within an octave are all approximate powers of the basic frequency ratio $1:2^{1/2}$, the semitone. Thus, there is some value of k for which

$$\left(\frac{3}{2}\right)^k \approx 2^{r/12} \qquad r = 1, 2, 3, \ldots$$

where k = r/7 is the solution of this approximate equation.

For the piano, the frequencies of the different keys are selected from the same basic set of frequencies. This has led to the development of Bach's tempered scales, based on the semitone ratio of $2^{1/12}$ (See Figure 1). The frequencies follow multiples of the lowest note:

$$1, 2^{1/12}, 2^{2/12}, 2^{3/12}, 2^{4/12}, \ldots$$

**Figure 1**

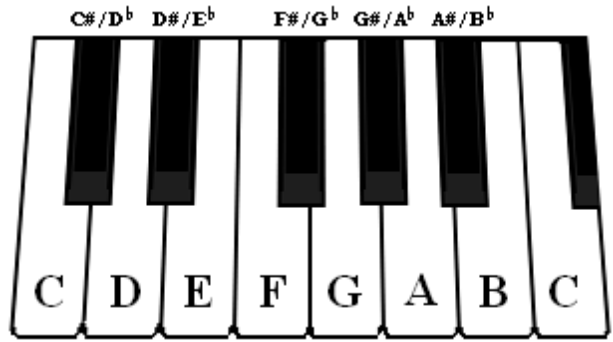

Ultimately, with the even-tempered system the sequence of pitch is not a precise agreement with the natural scale, but it does provide a close approximation. In fact, the modern ear (since the time of Bach in the 1700s) has become so accustomed to the errors that this tuning scheme sound correct.

**2. New Scale**

Some years ago J. R. Pierce[4] proposed to replace the frequency ratio 2:1 of the octave by the frequency ratio 3:1, which he called tritave. He designed an equal-tempered scale that matches frequency ratio constructed such that there is some integral root of 3, $3^{1/N}$, such that 5/3 and 7/3 are well approximated by integer powers of $3^{1/N}$. Analogously to the Bach's tempered scale, the frequency ratios form a self-similar sequence:

$$1, 3^{1/13}, 3^{2/13}, 3^{3/13}, 3^{4/13}, \ldots$$

As pointed out by Schroeder,[3] the number theory of this 13-step musical scale, so-called the Bohlen-Pierce (BP) scale, is nearly perfect, but its compositional value may still be open to debate. Although, the BP scale has gained great recognition in the literature, there are only rudimentary approaches available to a tonal understanding of it and instruments that can play it are few.

So far, existing synthesizer keyboards for the BP scale have been awkward adaptations of the regular piano scales. In some instances, one finds keyboards with covered and uncovered keys that make them unappealing for compositions.[5] Therefore, this work presents a new piano keyboard in order to make the BP scale more functional and pleasing for composition and performance.

### 3. Octave versus Tritave

For comparison, we find below the frequency correspondence between notes on the regular piano keyboard for two octaves in Figure 1. On the other hand, a tritave is the interval in this new scale from a note to the next note with the same letter. For example, the interval from one C to the next C in this scale is called a tritave. While the octave from the traditional scale is created by doubling the frequency, a tritave in this scale is created by tripling the frequency. The tuning of the entire scale is based on the thirteenth root of three, rather than the traditional twelfth root of two, which seems to be very mathematically promising. Within this scheme, a tritave "Major" scale can be expressed as follows: whole, whole, ½, whole, whole, ½, whole, ½.

Because of the added note, there is one extra white key, the H key, as can be seen in Figure 2 below.

**Figure 2**

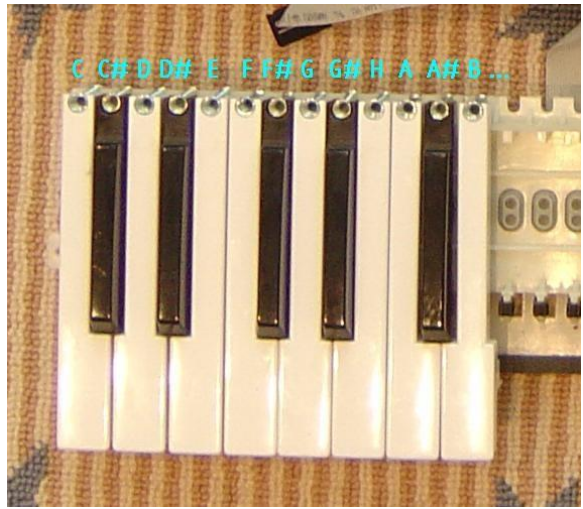

An audio clip of this song can be available on this website. You can listen to an MP3 (http://www.hwscience.com/HWJS/archives/keyboard/tritave.mp3) of the tritave scale. In this clip, the chromatic tritave scale (that is, all 13 notes from C to C) is played first, followed by the same chromatic scale with a root petal tone playing throughout to show all 13 possible harmonies. Some of these harmonies sound unpleasant (e.g., the C with the C#) and would obviously not be incorporated into any scale, just as some possible harmonies from the chromatic octave scale are not used either.

What is important with all these harmonies from the tritave scale is that many harmonies that are found in the octave scale (e.g., the major and minor third, fourth, fifth, octave, and more) also appear in the tritave scale (but in different spots), and, in addition, some new, very interesting harmonies can also be found in this scale (See Figure 3).

**Figure 3 (Tritave Scale)**

| Note | $C_2$ | $C_2^\#$ $D_2^b$ | $D_2$ | $D_2^\#$ $E_2^b$ | $E_2$ | $F_2$ | $F_2^\#$ $G_2^b$ | $G_2$ | $G_2^\#$ $H_2^b$ | $H_2$ | $A_3$ | $A_3^\#$ $B_3^b$ | $B_3$ | $C_3$ |
|---|---|---|---|---|---|---|---|---|---|---|---|---|---|---|
| Semitone Order in Octave | | 1 | 2 | 3 | 4 | 5 | 6 | 7 | 8 | 9 | 10 | 11 | 12 | 13 |
| Interval Ratio to Key Note | 1 | a | $a^2$ | $a^3$ | $a^4$ | $a^5$ | $a^6$ | $a^7$ | $a^8$ | $a^9$ | $a^{10}$ | $a^{11}$ | $a^{12}$ | $a^{13}$ |
| Frequency Ratio to Key Note | 1.000 | 1.088 | 1.184 | 1.289 | 1.402 | 1.526 | 1.660 | 1.807 | 1.966 | 2.139 | 2.328 | 2.533 | 2.757 | 3.000 |
| Frequency of Note (Hz) | 65.406 | 71.162 | 77.441 | 84.308 | 91.699 | 99.810 | 108.574 | 118.189 | 128.588 | 139.903 | 152.265 | 165.673 | 180.324 | 196.218 |

For comparison, the standard Equal-Tempered frequencies are included below.

**Figure 4 (Equal Tempered Scale)**

| Note | $C_2$ | $C_2^\#$ $D_2^b$ | $D_2$ | $D_2^\#$ $E_2^b$ | $E_2$ | $F_2$ | $F_2^\#$ $G_2^b$ | $G_2$ | $G_2^\#$ $A_3^b$ | $A_3$ | $A_3^\#$ $B_3^b$ | $B_3$ | $C_3$ |
|---|---|---|---|---|---|---|---|---|---|---|---|---|---|
| Semitone Order in Tritave | | 1 | 2 | 3 | 4 | 5 | 6 | 7 | 8 | 9 | 10 | 11 | 12 |
| Interval Ratio to Key Note | 1 | a | $a^2$ | $a^3$ | $a^4$ | $a^5$ | $a^6$ | $a^7$ | $a^8$ | $a^9$ | $a^{10}$ | $a^{11}$ | $a^{12}$ |
| Frequency Ratio to Key Note | 1.000 | 1.059 | 1.122 | 1.189 | 1.260 | 1.335 | 1.414 | 1.498 | 1.587 | 1.682 | 1.782 | 1.888 | 2.000 |
| Frequency of Note (Hz) | 65.406 | 71.162 | 77.441 | 84.308 | 91.699 | 99.810 | 108.574 | 118.189 | 128.588 | 139.903 | 152.265 | 165.673 | 180.324 |

## Acknowledgment

I would like to thank Scott Layne and David Katzenberg for their invaluable participation and interest in the lectures that led to this work

## References

1. J. R. Pierce, *The Science of Musical Sound*, Scientific American Books, Inc., N.Y., 1983.
2. H. F. Olson, *Music, Physics, and Engineering*, Dover Publication, Inc., N.Y., 1967.
3. M. Schroeder, *Fractals, Chaos, Power Laws*, W. H. Freeman and Company, N.Y., 1991.
4. M. V. Mathews, J. R. Pierce, A. Reeves, and L. Roberts, *Theoretical and Experimental Explorations of the Bohlen-Pierce Scale*, Journal of the Acoustical Society of America, 84, 1214-1222, 1988.
5. The Bohlen-Pierce site: http://members.aol.com/bpsie/index.html